\documentclass{MYappolb}            
\usepackage{epsfig}

\begin{document}
\title{The Aligned Two-Higgs-Doublet Model%
\thanks{Presented at the Flavianet topical workshop on {\it Low energy constraints on extensions of
the Standard Model}, Kazimierz, Poland, 23-27 July 2009}
}
\author{Paula Tuz\'{o}n
\and
Antonio Pich
\address{Departament de F\'{\i}sica Te\`{o}rica, IFIC, Universitat de Val\`{e}ncia - CSIC\\ Apt. Correus 22085, E-46071 Val\`{e}ncia, Spain}
}
\maketitle
\begin{abstract}
In the two-Higgs-Doublet model the alignment of the Yukawa flavour matrices of the two scalar doublets guarantees the absence of tree-level flavour-changing neutral couplings.
The resulting fermion-scalar interactions are parameterized in terms of three complex parameters $\zeta_f$,
leading to a generic Yukawa structure which contains as particular cases all known specific implementations of the model based on $\mathcal Z_2$ symmetries. These three complex parameters are potential new sources of CP violation.
\end{abstract}
\PACS{12.60.Fr, 12.15.-y, 12.15.Ji, 14.80.Cp}

\section{The generic two-Higgs-doublet model}
The two-Higgs-doublet model (THDM) \cite{guide, branco} is one of the most simplest extensions of the Standard Model (SM) and brings many interesting phenomenological features (new sources of CP violation, dark matter candidates, axion phenomenology \ldots).
 It is a gauge theory with the same fermion content as the SM and two Higgs doublets $\phi_a$ ($a=1,\, 2$) with hypercharge $Y=\frac{1}{2}$.
 The neutral components of the scalar doublets acquire vacuum expectation values that are, in general, complex: $<0|\phi_a^T(x)|0>=\frac{1}{\sqrt{2}}\, (0,v_a e^{i\theta_a})$. One can fix $\theta_1=0$ through an appropriate $U(1)_Y$ transformation, leaving
 the relative phase $\theta \equiv \theta_2 - \theta_1$.

A global SU(2) transformation in the scalar space $(\phi_1,\phi_2)$ allows us to define the so-called Higgs basis $(\Phi_1,\Phi_2)$ where only one doublet acquires a vacuum expectation value $v\equiv \sqrt{v_1^2+v_2^2}$\ :
\begin{eqnarray}
\left( \begin{array}{c} \Phi_1 \\ -\Phi_2  \end{array} \right) \equiv \; \frac{1}{v} \left[ \begin{array}{cc} v_1 & v_2 \\ v_2 & -v_1 \end{array} \right] \; \left( \begin{array}{c} \phi_1 \\ e^{-i\theta}\phi_2  \end{array} \right) \; ,
\end{eqnarray}
\begin{eqnarray}
\Phi_1=\left[ \begin{array}{c} G^+ \\ \frac{1}{\sqrt{2}}(v+S_1+iG^0) \end{array} \right] \; , \qquad\qquad
\Phi_2 = \left[ \begin{array}{c} H^+ \\ \frac{1}{\sqrt{2}}(S_2+iS_3)   \end{array}\right] \; .
\end{eqnarray}
In addition to the three Goldstone fields $G^\pm(x)$ and $G^0(x)$, there are
five physical degrees of freedom in the scalar sector: two charged fields $H^{\pm}(x)$ and three neutrals $\varphi_i^0(x)=\{h(x),H(x),A(x)\}$, which are related to the $S_i$ fields through an orthogonal transformation $\varphi^0_i(x)=\mathcal R_{ij}S_j(x)$. The form of $\mathcal R_{ij}$ depends on the scalar potential, which could violate CP in its most general version; in this case the resulting mass eigenstates
will not have a definite CP parity. If it is CP symmetric, the CP admixture disappears.

\subsection{Yukawa sector}

The most general Yukawa Lagrangian is given by
\begin{eqnarray}
\mathcal L_Y&=& \mbox{}-\bar{Q}_L' (\Gamma_1 \phi_1 +\Gamma_2 \phi_2)d_R'\, -\, \bar{Q}_L' (\Delta_1 \widetilde{\phi}_1 +\Delta_2 \widetilde{\phi}_2)u_R'
\nonumber \\  &&\mbox{}-
\bar{L}_L' (\Pi_1 \tilde{\phi}_1 + \Pi_2 \tilde{\phi}_2)l_R' \; +\; \mathrm{h.c.}  \; ,
\end{eqnarray}
where $ \bar{Q}_L'$ and $\bar{L}_L' $ are the left-handed quark and lepton doublets, respectively,
and  $\tilde{\phi_a}(x)\equiv i \tau_2\phi_a^*$ the charge-conjugated scalar fields.
All fermionic fields are written as $N_G$--dimensional vectors and the couplings $\Gamma_a$, $\Delta_a$ and $\Pi_a$ are $N_G\times N_G$ complex matrices in flavour space. Moving to the Higgs basis, the Lagrangian reads
\begin{eqnarray}
\mathcal L_Y &=& \mbox{} -\frac{\sqrt{2}}{v} \left\{  \bar{Q}_L'  (M_d'  \Phi_1 + Y_d' \Phi_2) d_R'\, +\, \bar{Q}_L'  (M_u' \tilde{\Phi}_1  + Y_u' \tilde{\Phi}_2)u_R'  \right.
\nonumber \\  &&\left. \hskip 1cm
\mbox{} + \bar{L}_L'  (M_l' \tilde{\Phi}_1  + Y_l' \tilde{\Phi}_2)l_R'\; +\; \mathrm{h.c.}  \right\} \; .
\end{eqnarray}
Here, $M_f'$ ($f=u,d,l$) are the non-diagonal fermion mass matrices and $Y_f'$ contain the Yukawa couplings to the scalar doublet that doesn't acquire a vacuum expectation value. Each right-handed fermion couples to two different 
matrices that in general cannot be diagonalized simultaneously. In the fermion mass-eigenstate basis $f(x)$, the $M_f'$ matrices become the diagonal mass matrices $M_f$, while $Y_f'$ become $Y_f$ which remain non-diagonal and unrelated to $M_f$. This originates dangerous flavour-changing neutral-current (FCNC) interactions, which are tightly constrained phenomenologically.

To avoid this problem one can assume that the non-diagonal Yukawa couplings are proportional to the geometric mean of the two fermion masses, $g_{ij}\propto \sqrt{m_im_j}$ \cite{cheng}. This kind of phenomenologically viable models, known as Type III THDM \cite{atwood},
can be obtained assuming particular textures of the Yukawa matrices \cite{cheng}. Another possibility is to make the scalars heavy enough to suppress low-energy FCNC effects. However, this assumption leads to a phenomenologically non relevant THDM.

 A more elegant solution is to impose a discrete $\mathcal Z_2$ symmetry to enforce that only one scalar doublet couples to a given right-handed fermion \cite{glashow}; one of the two Yukawa matrices has to be zero. One requires
 the Lagrangian to remain invariant under the change $\phi_1 \rightarrow \phi_1$, $\phi_2 \rightarrow -\phi_2$, $Q_L \rightarrow Q_L$, $L_L \rightarrow L_L$, and appropriate transformation properties for the right-handed fermions. There are four non-equivalent 
 choices: the Type I (only $\phi_2$ couples to fermions) \cite{haber, hall, savage}, Type II ($\phi_1$ couples to $d$ and $l$ and $\phi_2$ to $u$) \cite{hall, donoghue, savage}, Type X or leptophilic ($\phi_1$ couples to fermions and $\phi_2$ to quarks) and Type Y ($\phi_1$ couples to $d$ and $\phi_2$ to $u$ and $l$) models \cite{savage, barger, grossman, akeroyd, aoki}.
 The $\mathcal Z_2$ symmetry can be also implemented in the Higgs basis, forcing all fermions to couple to $\Phi_1$; this \emph{inert} doublet model is a natural frame for dark matter \cite{ma, barbieri, lopez}.

\section{Aligned THDM}
A more general way to avoid tree-level FCNCs is to require the alignment in flavour space of the Yukawa coupling matrices \cite{nos}:
\begin{eqnarray}
\Gamma_2 = \xi_d e^{-i\theta} \Gamma_1 \; , \qquad \Delta_2=\xi_u^* e^{i\theta}\Delta_1\; ,  \qquad \Pi_2=\xi_le^{-i\theta} \Pi_1 \; ,
\end{eqnarray}
with $\xi_f$ arbitrary complex numbers. The alignment guarantees that the matrices $Y'_f$ and $M'_f$ are proportional and can be simultaneously diagonalized:
\begin{eqnarray}
Y_{d,l} = \zeta_{d,l} M_{d,l}\; , \qquad Y_u = \zeta_u M_u \; ,  \qquad \zeta_f \equiv \frac{\xi_f-\tan{\beta}}{1+\xi_f\tan{\beta}} \; ,
\end{eqnarray}
where $\tan{\beta}\equiv v_2/v_1$.
In terms of mass-eigenstate fields, the Yukawa Lagrangian takes then the form:
\begin{eqnarray}
\mathcal L_Y &=& \mbox{} -\frac{\sqrt{2}}{v} H^+(x) \left\{  \bar{u}(x)   \left[  \zeta_d V M_d \mathcal P_R - \zeta_u M_u V \mathcal P_L  \right]   d(x) + \zeta_l \,\bar{\nu}(x) M_l \mathcal P_R l(x)\right\}
\nonumber \\ && \mbox{}  - \frac{1}{v} \,\sum_{\varphi, f} \varphi^0_i(x) y^{\varphi^0_i}_f \; \bar{f}(x)\;  M_f \mathcal P_R  f(x)
\; +\;  \mathrm{h.c.} \; ,
\end{eqnarray}
where $V$ is the CKM matrix, $\mathcal P_{R,L}\equiv \frac{1\pm \gamma_5}{2}$ and $y^{\varphi^0_i}_f$ are neutral couplings of the physical scalar fields \cite{nos}. Therefore, within this approach:
\begin{itemize}
\item[--] All scalar-fermion couplings are proportional to the corresponding fermion masses.
\item[--] The neutral Yukawas are diagonal in flavour.
\item[--] The only source of flavour-changing interactions is the quark-mixing matrix V in the charged sector.
\item[--] All leptonic couplings are diagonal in flavour, since there are no right-handed neutrinos in the theory.
\item[--] There are only three new parameters $\zeta_f$, which encode all possible freedom allowed by the alignment conditions. These couplings satisfy universality among the different generations, i.e. there is the same universal coupling for all fermions with a given electric charge. They are also invariant under global SU(2) transformations of the scalar fields $\phi_a \rightarrow \phi_a'=U_{ab}\phi_b$ \cite{davidson}, i.e. they are scalar basis independent.
\item[--] The usual models based on $\mathcal Z_2$ symmetries are recovered taking the appropriated limits shown at Table 1.
\item[--] The $\zeta_f$ can be arbitrary complex numbers. Their phases introduce new sources of CP violation without tree-level FCNCs.
\end{itemize}

\begin{table}[t]
\centering
\caption{Choices of $\xi_f$ associated with $\mathcal Z_2$ models and corresponding values of $\zeta_f$.}
\begin{tabular}{|c|c|c|c|c|}
\hline
Model & $(\xi_d,\xi_u,\xi_l)$ & $\zeta_d$ & $\zeta_u$ & $\zeta_l$  \\
\hline
Type I & $(\infty,\infty,\infty)$& $\cot{\beta}$ &$\cot{\beta}$ & $\cot{\beta}$ \\
Type II & $(0,\infty,0)$& $-\tan{\beta}$ & $\cot{\beta}$ & $-\tan{\beta}$ \\
Type X & $(\infty,\infty,0)$& $\cot{\beta}$ & $\cot{\beta}$ & $-\tan{\beta}$ \\
Type Y & $(0,\infty,\infty)$& $-\tan{\beta}$ & $\cot{\beta}$ & $\cot{\beta}$ \\
Inert & $(\tan{\beta},\tan{\beta},\tan{\beta})$ & 0 & 0 & 0 \\
\hline
\end{tabular}
\end{table}

Lepton-flavour-violating neutral couplings are identically zero to all orders in perturbation theory, because of the absence of right-handed neutrinos. So, the usually adopted $\mathcal Z_2$ symmetries are not necessary in the lepton sector, making the Type X and Y models less compelling.
From a phenomenological point of view, the coupling $\zeta_l$ could take any value.

Quantum corrections could introduce some misalignment of the quark Yukawa coupling matrices, generating small FCNCs suppressed by the corresponding loop factors. However, the flavour mixing induced by loop corrections has a very characteristic structure \cite{nos} which resembles the popular Minimal Flavour Violation scenarios \cite{chivukula}. This restriction in the structure of the local FCNC terms is due to the following symmetry of the aligned THDM: the Lagrangian remains invariant under flavour-dependent phase transformations of the different fermion mass eigenstates, $f_i(x) \rightarrow e^{i\alpha^f_i}f_i(x)$, provided the quark mixing matrix is transformed as $V_{ij}\rightarrow e^{i\alpha^u_i}V_{ij} e^{-\alpha^d_j}$
\cite{nos}. The only possible FCNC terms are then of the type \ $\bar u_i [V(M_d)^nV^{\dagger}]_{ij} u_j$ \ and \ $\bar d_i [V^{\dagger}(M_u)^mV]_{j} d_j$ \ $(n,m>0)$, or similar structures with additional factors of $V$ and $V^{\dagger}$ \cite{nos}.
Structures of this type have been recently discussed in Ref.~\cite{bbr09}.

\section{Phenomenology}

One of the most distinctive features of the THDM is the presence of a charged scalar $H^{\pm}$.
Assuming that $\mathrm{Br}(H^+\rightarrow c \bar{s})+\mathrm{Br}(H^+\rightarrow \tau^+ \nu_{\tau})=1$, the combined LEP data
on $e^+e^- \rightarrow H^+ H^-$ constrain $M_{H^{\pm}}>78.6$ GeV (95\% CL) \cite{aleph}.
The CDF \cite{cdf} and D0 \cite{d0} collaborations have searched for $t\rightarrow H^+ b$ decays with negative results. CDF assumes $\mathrm{Br}(H^+\rightarrow c\bar{s})=1$ ($\zeta_l<<\zeta_{u,d}$), while D0 adopts the opposite hypothesis $\mathrm{Br}(H^+\rightarrow \tau \nu_{\tau})=1$ ($\zeta_l>>\zeta_{u,d}$). Both experiments find upper bounds on $\mathrm{Br}(t\rightarrow H^+ b )$ around 0.2 ($95\%$ CL) for charged scalar masses between 60 and 155 GeV. This implies $|\zeta_u|<0.3-2.5$, where the exact number depends on $M_{H^{\pm}}$.

One can also extract information from the semileptonic decays of a pseudoscalar meson $P^+\rightarrow l^+ \nu_l$, which are sensitive to $H^+$ exchange due to the helicity suppression of the SM amplitude: $\Gamma(P^+_{ij}\rightarrow l^+ \nu_l)/\Gamma(P^+_{ij}\rightarrow l^+ \nu_l)_{SM}=|1-\Delta_{ij}|^2$, where the new physics information is encoded in $\Delta_{ij}=(m_{P^{\pm}_{ij}}/M_{H^{\pm}})^2\zeta_l^*(\zeta_um_{u_i}+\zeta_dm_{d_j})/(m_{u_i}+m_{d_j})$. The correction $\Delta_{ij}$ is in general complex and its real part can have either sign. To determine its size one needs to know $V_{ij}$ and a theoretical determination of the meson decay constant. From the present knowledge of $B^+\rightarrow \tau^+ \nu_{\tau}$ \cite{antonelli}, we obtain $|1-\Delta_{ub}|=1.32\pm 0.20$ which implies $M_{H^{\pm}}/\sqrt{|Re(\zeta_l^+\zeta_d)|}>5.7$ GeV ($90\%$ CL). With $f_{D_s}=242\pm6$ MeV \cite{follana}, the recent CLEO measurement $\mathrm{Br}(D_s^+\rightarrow \tau^+ \nu_{\tau})=(5.62\pm0.64)\%$ \cite{cleo} implies $|1-\Delta_{cs}|=1.07\pm0.05$ and $M_{H^{\pm}}/\sqrt{|Re(\zeta_l^+\zeta_u)|}>4.4$ GeV ($90\%$ CL).

FCNC processes such as $b\rightarrow s \gamma$ or $P^0-\bar{P}^0$ mixing are also very sensitive to scalar contributions and give strong constraints on the parameter space $(M_{H^{\pm}}, \zeta_{u,d})$. A detailed phenomenological analysis will be presented somewhere else.

\section*{Acknowledgements}
This work has been supported in part by
the EU Contract MRTN-CT-2006-035482 (FLAVIAnet),
by MICINN, Spain [grants FPA2007-60323 and Consolider-Ingenio 2010 CSD2007-00042 –CPAN–]
and by Generalitat Valenciana (PROMETEO/2008/069). P.T is indebted to the Spanish MICINN for a FPU
fellowship.

\end{document}